# Transition Metal-Tetracyanoquinodimethane Monolayers as Single-Atom Catalysts for Electrocatalytic Nitrogen Reduction Reaction

*Yiran Ying [a], Ke Fan [a], Xin Luo [b]\*, and Haitao Huang [a]\**

[a] Department of Applied Physics, The Hong Kong Polytechnic University, Hung Hom, Kowloon, Hong Kong, P.R. China

[b] School of Physics, Sun Yat-sen University, Guangzhou, Guangdong Province, P.R. China, 510275

AUTHOR INFORMATION

**Corresponding Author**

*E-mail: luox77@mail.sysu.edu.cn (X.L.); aphhuang@polyu.edu.hk (H.H.).

ABSTRACT.

Converting earth-abundant dinitrogen into value-added chemical ammonia is a significant yet challenging topic. Electrocatalytic nitrogen reduction reaction (NRR), compared with conventional Haber-Bosch process, is an energy-saving and environmentally friendly approach. The major task of electrocatalytic NRR is to find electrocatalysts which can activate dinitrogen effectively and exhibit high selectivity and stability. Single atom catalysts can act as a good solution. In this work, by means of first-principles density functional theory, molecular dynamics calculations, and a two-step screening process, we confirm that single Sc and Ti atom supported on tetracyanoquinodimethane monolayers (Sc/Ti-TCNQ) are excellent candidates for NRR electrocatalysts. $N_2$ adsorption and activation are effective due to the 'acceptance-donation' mechanism and outstanding electronic structure of TM-TCNQ, and Gibbs free energy diagram shows that Sc-TCNQ and Ti-TCNQ exhibit low NRR overpotential of 0.33 and 0.22 V through enzymatic-consecutive mixed pathway, respectively. In addition, selectivity over HER and stability of Sc/Ti-TCNQ monolayers are also validated. This work opens a new avenue for designing novel single atom catalysts for NRR as well as other catalytic applications.

# 1. Introduction

Ammonia ($NH_3$) is an indispensable chemical in industry and agriculture because it not only can serve as a precursor to the production of many kinds of fertilizers, chemicals for daily use, and pharmaceuticals, but also is a clean, carbon-free energy carrier whose combustion products (dinitrogen and water) are all environmentally benign.[1-3] Nitrogen fixation, converting dinitrogen ($N_2$)—the most abundant gas in Earths' atmosphere which cannot be directly utilized by human beings—into value-added chemical ammonia, is undeniable an urgent topic ever since human industrialization. According to Mineral Commodity Summaries from US Geological Survey, about 170 million tons of ammonia in total were produced globally in 2018.[4] A major challenge in nitrogen fixation is breaking the chemically inert N≡N triple bond, leading to the fact that conventional nitrogen fixation approach—Haber-Bosch (H-B) process requires high temperature and pressure and is highly energy consuming, accounting for around 2% of global supply of energy.[5-10] Besides, H-B process needs large plant infrastructure and may raise environmental issues such as emission of carbon dioxide, taking up about 1% of total greenhouse gas emission.[10, 11]

Electrocatalytic nitrogen reduction reaction (NRR, $N_2+6H^++6e^-->2NH_3$), inspired by the biological NRR with nitrogenase enzymes, can be a promising alternative for H-B process because the reaction can take place at ambient condition.[7, 11, 12] Discovering the electrocatalysts for NRR with high activity and selectivity is a significant task for both fundamental research and industrial applications.[13] The past decade has witnessed the efforts to investigating metals,[13, 14] metal oxides,[15] metal nitrides,[16] metal carbides,[17] and metal phosphides[18] as NRR electrocatalysts by means of density functional theory (DFT) calculations. However, the relatively low utilization percentage of active materials and high cost of these metal-based electrocatalysts hinder their experimental realization.

Recently, another group of catalysts—single-atom catalysts (SACs), have appeared in the vision of researchers. SACs, which are defined as isolated atoms anchored on the substrates, exhibit remarkable catalytic performance and relatively lower cost because of the full utilization of active metal atoms, high selectivity, and stability.[19] Aside from these features, transition metal (TM) based SACs can be very effective especially for NRR because $d$ electrons of the single metal supported on substrate are more active than those in the aggregated form, and are more possible to activate the N≡N triple bond. DFT calculations predict that transition metal (TM) atoms embedded in two-dimensional boron nitride,[20] nitrogen-doped graphene,[21,22] graphitic carbon nitride,[23-25] and transition metal dichalcogenides,[26,27] are outstanding SACs for NRR. Nevertheless, in experiments, examples of SACs for NRR are scarce and only limited to single transition metal (TM) supported on nitrogen-doped carbon[28-31] and more efforts into this burgeoning area are still needed.

Metal-organic frameworks (MOFs) have recently demonstrated their potential to form SACs due to tunable geometric structure and large surface area.[32-34] 7,7,8,8-tetracyanoquinodimethane (TCNQ), an outstanding electron acceptor, can serve as a substrate for forming MOFs. Many transition metal supported on TCNQ structures have been synthesized in experiments with their electronic and magnetic properties widely studied.[35-37] For electrocatalytic applications, on the other hand, different TM-TCNQ monolayers have been predicted by DFT calculations to be promising electrocatalysts for oxygen evolution and reduction[38,39], CO oxidation[40], and $CO_2$ reduction[41]. However, the applications of TM-TCNQ as electrocatalysts for NRR have not been investigated theoretically or experimentally.

Based on the information given above, in this work, we use DFT calculations to search for the potential candidates as electrocatalysts for NRR in TM-TCNQ monolayers by screening 17 TM elements which are common on the earth (Sc-Zn, Mo, Ru-Pd, Ag, Pt, Au). Gibbs free energy calculations are conducted to evaluate the NRR performance of the selected samples, and stability and NRR selectivity of these SACs are further checked.

## 2. Computational Methods

All spin-polarized DFT and AIMD calculations were performed by employing the projector-augmented wave method implemented in the Vienna *ab initio* Simulation Package (VASP).[42, 43] Perdew-Burke-Ernzerhof (PBE) flavor of the generalized gradient approximation[44] was chosen as the exchange-correlation functional. For the consideration of van der Waals interactions, Grimme's DFT-D3 semiempirical scheme was applied.[45] Kinetic energy cut-off was set as 400 eV and the first Brillouin zone was sampled with Gamma-centered 7×5×1 Monkhorst-Pack meshes. Both cell and ionic degrees of freedom of all structures were allowed to relax until convergence criteria ($10^{-5}$ eV for energy and 0.02 eV Å$^{-1}$ for force) were satisfied. Vacuum layer with thickness larger than 20 Å was used to avoid spurious interactions between adjacent cells. Grid-based Bader charge analysis[46] was used to calculate the charge transfer. AIMD simulations were performed in NVT ensemble with Nosé-Hoover thermostat.[47, 48] Structures were visualized with VESTA package[49], and VASPKIT code[50] was used for processing the calculation results from VASP.

NRR was considered as a six net coupled proton and electron transfer (CPET) process and one coupled proton and electron was transferred from the solution to the surface of electrocatalyst for each step. The Gibbs free energy was calculated using the following formula:[51]

$$\Delta G = \Delta E_{DFT} + \Delta E_{ZPE} - T\Delta S + \Delta G_{pH}$$

where $\Delta E_{DFT}$, $\Delta E_{ZPE}$, $T\Delta S$, and $\Delta G_{pH}$ represent the difference of DFT-calculated adsorption energy, zero point energy (ZPE), entropic contribution, and free energy correction of pH in each reaction step. Conventional hydrogen electrode (CHE) model to calculate the Gibbs free energy of $(H^+ + e^-)$:[52]

$$G(H^+ + e^-) = 0.5 G(H_2) - eU$$

where U represents the electrode potential relative to the CHE. ZPE and entropic contribution were calculated from vibrational frequencies $\omega_i$:[17, 53, 54]

$$E_{ZPE} = \frac{1}{2} \sum_i \hbar \omega_i$$

$$TS = RT \left\{ \sum_i \frac{\frac{\hbar \omega_i}{k_B T}}{\exp\left(\frac{\hbar \omega_i}{k_B T}\right) - 1} - \sum_i \ln[1 - \exp\left(-\frac{\hbar \omega_i}{k_B T}\right)] \right\}$$

where $k_B$ and R represent Boltzmann constant and gas constant, respectively.

# 3. Results and Discussions

## 3.1 Screening TM-TCNQ Monolayers for NRR and the $N_2$ activation

The structures of TM-TCNQ are modelled by 2×2 supercell with twelve carbon atoms, four nitrogen atoms, four hydrogen atoms, and one TM atom. A prototypical DFT-optimized structure of TM-TCNQ—Sc-TCNQ is shown in Fig. 1. The lattice constants for TM=Sc, Ti, V, Cr, Mn, Fe, Co, Ni, Cu, Zn, Mo, Ru, Rh, Pd, Ag, Pt, and Au are listed in Table S1.[41] All TM-TCNQ structures exhibit planar structures and each TM atom coordinates with four nitrogen atoms from TCNQ. From Bader charge analysis, in Sc-TCNQ, Ti-TCNQ, and V-TCNQ, TM atom transfers 1.87 $e^-$, 1.80 $e^-$, and 1.46 $e^-$ to the TCNQ substrate, indicating that TCNQ can accept electrons from TM atoms. We further calculate the electron localized function (ELF) and show the contour plot in Fig. 1a. ELF values close to 0.5 and 1 suggest metal bond (delocalized electrons) and covalent bond (localized electrons), respectively, while ELF values smaller than 0.5 represent ionic bond.[55, 56] From Fig. 1a, we can conclude that ionic bond forms between Sc and adjacent N atoms. Inside TCNQ substrate, on the other hand, covalent bond forms between atoms.

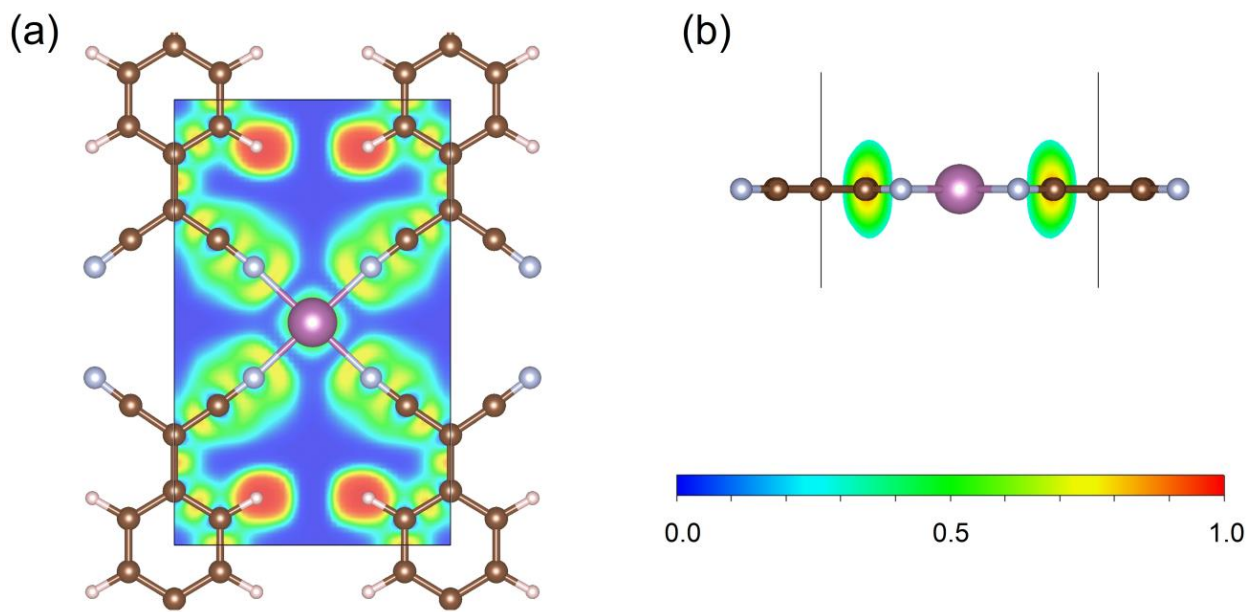

**Fig. 1** (a) Top and (b) side view of the optimized structure of Sc-TCNQ and corresponding electron localized function contour map. Sc, C, N, and H atoms are represented in purple, brown, cyan, and pale pink, respectively. Black line denotes the cell boundary of Sc-TCNQ. Isodensity values for ELF map are shown below Fig.1b.

The $N_2$ adsorption and activation of the N≡N triple bond is the first and vital step for NRR.[12, 13] In this concern, before proceeding to study NRR reaction pathways, we first examine the $N_2$ adsorption on TM-TCNQ. Two adsorption patterns of $N_2$ are considered—side-on configuration with both N atoms bind with TM atom (active site for NRR), and end-on configuration with one of the N atoms bind with TM atom (Fig. S1). Each configuration could lead to different NRR reaction pathways. From pioneer theoretical investigations, NRR is a complicated process and several pathways have been concluded.[20, 22, 57] For end-on configuration, distal, alternating, and their mixed mechanisms can be possible (Fig. S2); for side-on configuration, enzymatic, consecutive, and their mixed mechanisms are proposed (Fig. 2). As a result, screening the $N_2$ adsorption ability and configuration on TM-TCNQ for each TM is our top priority.

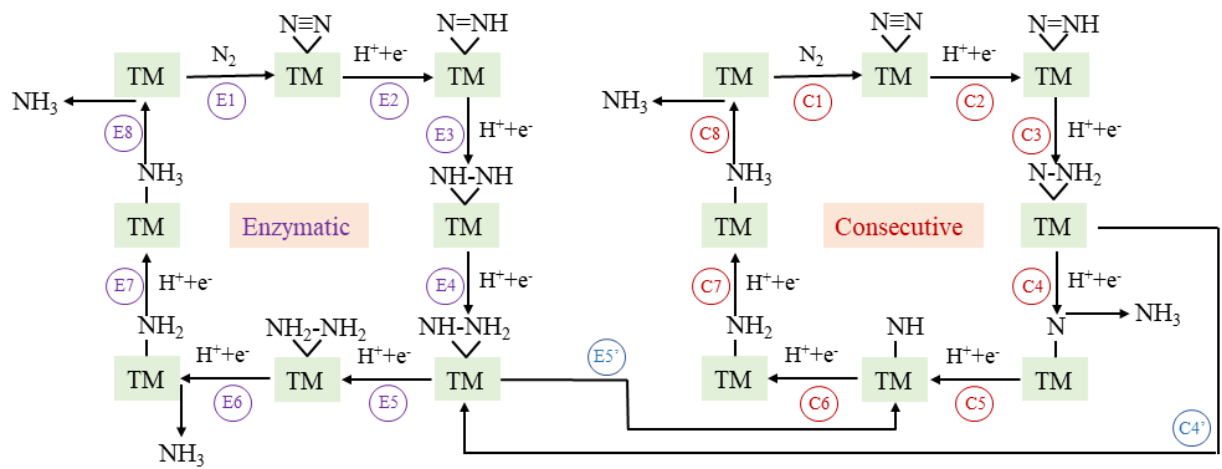

**Fig. 2** Schematic illustration of enzymatic, consecutive, and their mixed mechanism for NRR on TM-TCNQ. TM denotes transition metal atom (active site) in the figures.

$N_2$ adsorption energy values $E_{ad}$, defined as $E_{ad}=E(TM\text{-}TCNQ+N_2)-E(TM\text{-}TCNQ)-E(N_2)$, are calculated and summarized in Table S2 and Fig. 3(a) for both side-on and end-on pattern on TM-TCNQ. For effective activation of $N_2$, $E_{ad}$ should be negative enough (<-0.3 eV) to ensure the energetically favorable $N_2$ adsorption. From Fig. 3(a), six cases of end-on configuration: Sc-TCNQ, Ti-TCNQ, V-TCNQ, Cr-TCNQ, Mn-TCNQ, Fe-TNCQ, and three cases of side-on configuration: Sc-TCNQ, Ti-TCNQ, V-TCNQ are selected from a total of 34 structures (Fig. 3a). It is worth noticing that even though side-on configurations of $N_2$ on Sc, Ti, V-TCNQ are less energetically favorable than the corresponding end-on configurations, their $E_{ad}$ values can still meet the criteria of -0.3 eV.

To further screening the remaining nine structures for NRR, we further calculate the Gibbs free energy change for *N≡N + (H$^+$+e$^-$) -> *N=NH (end-on) and *N≡N* + (H$^+$+e$^-$) -> *N=NH* (side-on), because these steps (the first hydrogenation step in Fig. 2 and Fig. S2) are proven to be the potential limiting step (PLS) for multiple cases of TM-based SACs for NRR from previous publications.[21, 58-61] Results shown in Fig. 3b indicate that for all six end-on structures, the corresponding Gibbs free energy change values ΔG(NNH) are larger than 1.0 eV. These values are much higher than most of the reported SACs for NRR and will lead to high overpotential and mediocre NRR performance. On the other hand, for Sc, Ti, V-TCNQ, side-on N$_2$ adsorption patterns can result in ΔG(NNH) values smaller than 1.0 eV. To conclude, after the two-step screening process shown above, we have identified these 3 structures out of 34 for further investigations.

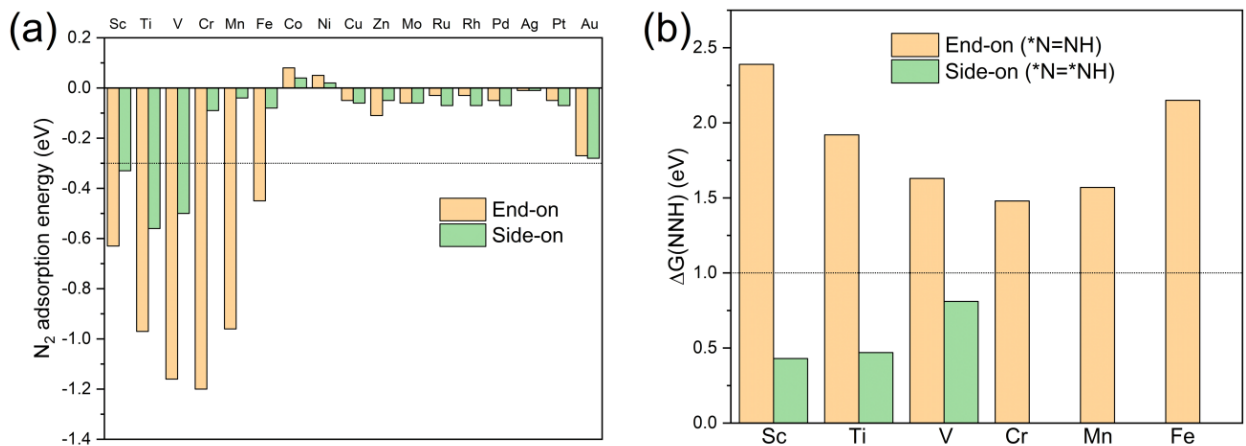

**Fig. 3** The two-step screening of TM-TCNQ electrocatalysts for NRR. (a) Calculated N$_2$ adsorption energy for both end-on and side-on configuration on TM-TCNQ (TM=Sc-Zn, Mo, Ru-Pd, Ag, Pt, Au), and (b) Gibbs free energy change values ΔG(NNH) of *N≡N + (H$^+$+e$^-$) -> *N=NH for Sc, Ti, V, Cr, Mn, Fe-TCNQ and *N≡N* + (H$^+$+e$^-$) -> *N=NH* for Sc, Ti, V-TCNQ.

Electronic structure is an important factor for evaluating the $N_2$ activation on NRR catalysts. Use Sc-TCNQ as an example, we find that density of states (DOS) for pure Sc-TCNQ (Fig. 4a) clearly show the metallic character, which facilitates fast charge transfer and is beneficial to the potential electrocatalytic process. DOS plot for $N_2$ in gas phase is show in Fig. 4b, where the lowest unoccupied molecular orbital (LUMO) and the highest occupied molecular orbital (HOMO) are respectively located at $\pi_g^*2p$ and $\sigma_g2p$ orbitals, which are consistent with literature.[62] $\pi_g^*2p$, $\sigma_g2p$, and $\pi_\mu2p$ orbitals from $N_2$ all match with DOS for Sc-TCNQ, leading to the hybridization and delocalization of orbitals in Sc-TCNQ with $N_2$ adsorption (Fig. 4c). This can lead to the effective activation of $N_2$ on the catalyst, which can be confirmed by the fact that N-N bond is elongated from 1.098 Å (in gas phase $N_2$) to 1.16 Å.

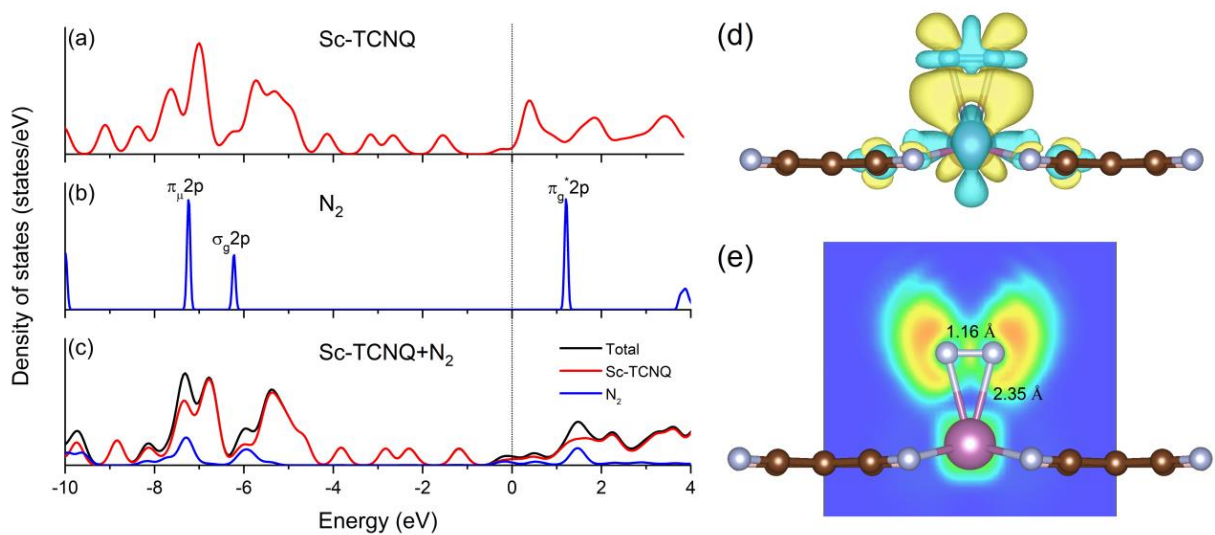

**Fig. 4** Density of states for (a) pristine Sc-TCNQ, (b) $N_2$ molecule in the gas phase, and (c) Sc-TCNQ with one adsorbed $N_2$ molecule. Fermi level is set to zero and denoted in black dotted line. (d) Charge density difference distribution (charge accumulation and depletion are shown in yellow

and light blue) of N$_2$ adsorption with the side-on configuration, and (e) corresponding ELF map with bond length of Sc-N and N-N marked in the figure.

Sc, Ti, and V are all early transition metal atoms with both occupied and unoccupied d orbitals, and the corresponding TM-based catalysts are promising for NRR because of the 'acceptance-donation' mechanism.[54, 63] Unoccupied TM d orbitals can accept lone-pair electrons from N$_2$, and in the meantime, occupied TM d electrons can be donated to the antibonding orbitals of N$_2$, leading to the strong binding of N$_2$. To unravel the electron transfer during the adsorption of N$_2$, we calculate the charge density difference distribution as $\Delta\rho=\rho(\text{Sc-TCNQ}+\text{N}_2)-\rho(\text{Sc-TCNQ})-\rho(\text{N}_2)$, where $\rho$ denotes the charge density distribution. Results in Fig. 4d exhibit that both charge accumulation (yellow) and depletion (light blue) happen around the adsorbed N$_2$ and Sc atom (active site for NRR), confirming that 'acceptance-donation' process happens in Sc-TCNQ. Charge transfer is also validated by Bader charge analysis result that Sc-TCNQ transfers 0.32 e$^-$ to the adsorbed N$_2$ molecule. In addition, ELF map (Fig. 4e) shows that ionic bonds form between Sc and the adsorbed N.

### 3.2 Gibbs Free Energy Diagrams and NRR Reaction Pathways

To gain a deeper view of the most probable NRR mechanisms, we perform calculations of Gibbs free energy for each intermediate in each step (Table S3-S5) to obtain the Gibbs free energy diagram for NRR. Here we only focus on the enzymatic, consecutive, and the mixed mechanism (Fig. 2) for Sc, Ti, V-TCNQ after the two-step screening discussed above (Fig. 3) and the result is shown in Fig. 5. The DFT- optimized structures for the relevant intermediates for Sc and Ti-TCNQ are plotted in Fig. S3 as examples. In the calculations, six consecutive protonation steps are considered, and two ammonia molecules are formed with the general formula of $N_2+6(H^++e^-) \rightarrow 2NH_3$.

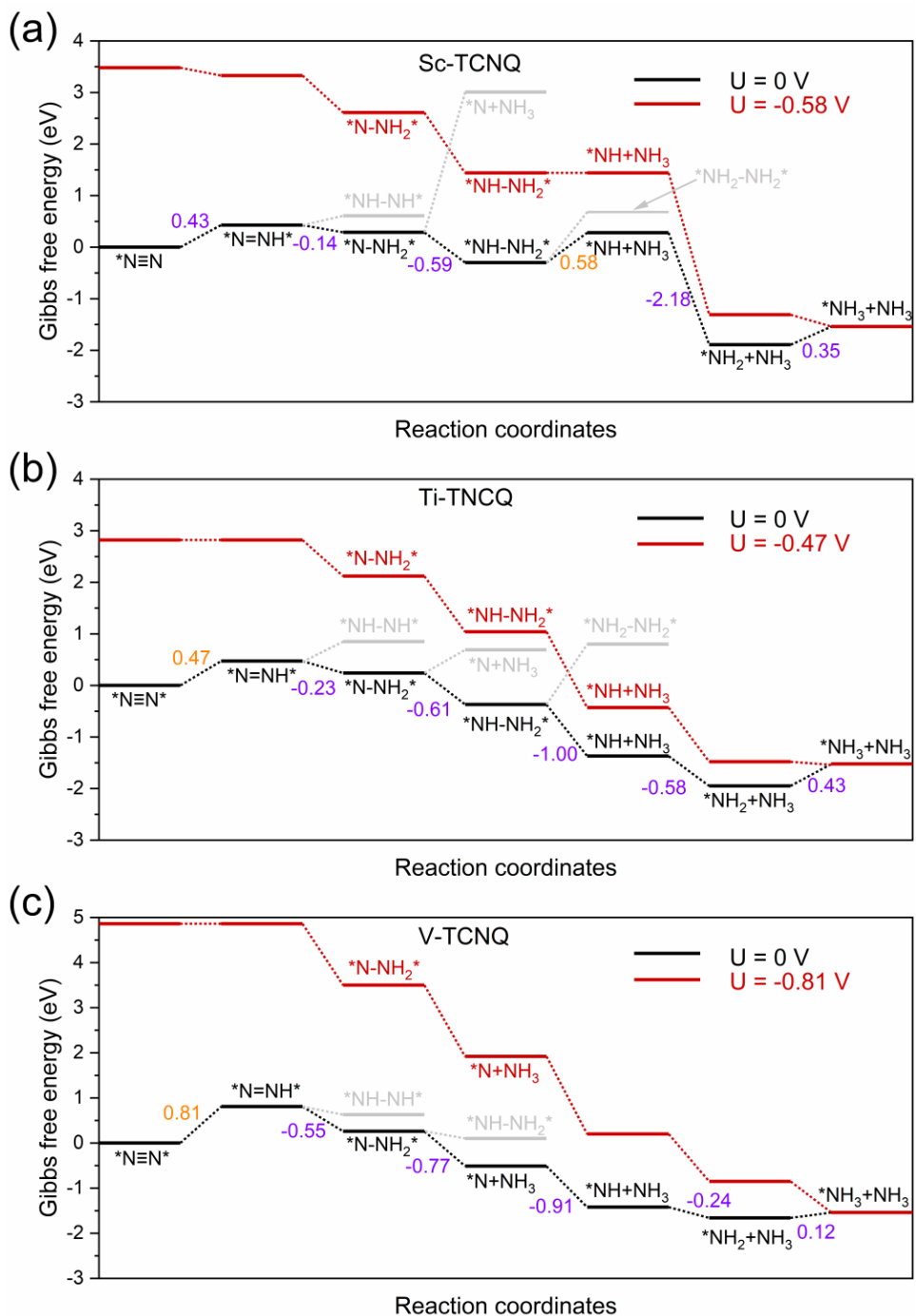

**Fig. 5** Gibbs free energy diagram for nitrogen reduction reaction through enzymatic/consecutive/mixed mechanism on (a) Sc-TCNQ, (b) Ti-TCNQ, and (c) V-TCNQ at zero and applied potential U. For U=0, reaction pathways and intermediates with unfavorable Gibbs free energy change are marked in grey, and the Gibbs free energy changes for each step are marked and shown in eV (value marked in orange color denotes potential limiting step).

For the first hydrogenation step *N≡N*-> *N=NH*, all three systems undergo a Gibbs free energy uphill of 0.43, 0.47, and 0.81 eV for Sc-TCNQ, Ti-TCNQ, and V-TCNQ, and the N-N bonds are elongated from 1.16, 1.17, and 1.17 Å in *N≡N* to 1.24, 1.24, and 1.23 Å in *N=NH*, respectively. Comparison of Gibbs free energy of *NH-NH* and *N-NH$_2$* determines whether the second step goes through consecutive or enzymatic pathway. From Fig. 5, free energy values of *NH-NH* are higher than those of *N-NH$_2$* for all systems, and *N=NH*->*N-NH$_2$* is an exothermic step. The N-N bond length is further elongated to over 1.4 Å. Different preferable pathways occur in the third step, where Sc and Ti-TCNQ prefer *N-NH$_2$*->*NH-NH$_2$* and V-TCNQ prefers *N-NH$_2$*->*N+NH$_3$ (one ammonia molecule released), but the step is still spontaneous. In the fourth protonation step, Sc and Ti-TCNQ tend to disengage one NH$_3$ molecule and form *NH. There is an energy uphill for Sc-TCNQ by 0.58 eV, while for Ti and V-TCNQ, the fourth step is exothermic. The following two steps *NH->*NH$_2$->*NH$_3$ and only the last step is endothermic with energy change of 0.35, 0.43, and 0.12 eV for Sc, Ti, and V-TCNQ, respectively. The energetically favorable pathway for Sc and Ti-TCNQ is enzymatic-consecutive mixed pathway, while for V-TCNQ, consecutive pathway is adopted.

Potential-limiting step (PLS) is defined as the reaction step with the highest Gibbs free energy increase and PLS determines the performance of the electrocatalysts. From Fig. 5, we can conclude that the PLS for Ti and V-TCNQ is the first protonation step *N≡N*-> *N=NH* with 0.47 and 0.81 eV free energy change, and for Sc-TCNQ, the fourth step *NH-NH$_2$*->*NH+NH$_3$ is PLS (0.58 eV). With a limiting potential U=-0.58, -0.47, and -0.81 V for Sc, Ti, and V-TCNQ, respectively, all reaction pathways can be exothermic (Fig. 5). Usually, overpotential (η) is an indicator for determining catalytic activity, and η for NRR is defined as η=U$_{equilibrium}$-U$_{limiting}$,[20] and U$_{equilibrium}$ is calculated to be -0.25 V in this work (Gibbs free energy change for the total reaction). Therefore, η values for Sc-TCNQ, Ti-TCNQ, and V-TCNQ are 0.33 V, 0.22 V, and 0.56 V, respectively. The overpotential values for Sc and Ti-TCNQ are much lower than the majority of reported NRR SACs, such as Mo on nitrogen doped graphene (N$_3$-G) (0.34 V)[21], Mo on graphene-boron nitride hybrid sheet (0.42 V)[64], and Fe on MoN$_2$ (0.47 V)[65], indicating that Sc-TCNQ and Ti-TCNQ are outstanding electrocatalysts for NRR.

To further understanding the NRR catalytic activity of TM-TCNQ, we perform Bader charge analysis along the most favorable reaction pathway, and each intermediate is divided into three moieties: the TCNQ nanosheet (denoted as moiety 1), single TM atom (denoted as moiety 2), and the adsorbed N$_x$H$_y$ (moiety 3). Results for Sc and Ti-TCNQ are plotted in Fig. 6. We can see that for all three moieties, there is obvious charge fluctuation, but generally, the TCNQ nanosheet act as an electron acceptor, receiving over 1.0 e$^-$ for each step while the TM atom donates more than 1.5 e$^-$. Nevertheless, the fluctuation of Bader charge for TM is very small, indicating that TM atom also acts as a transmitter to the charge transfer between the TCNQ nanosheet and N$_x$H$_y$ moiety.

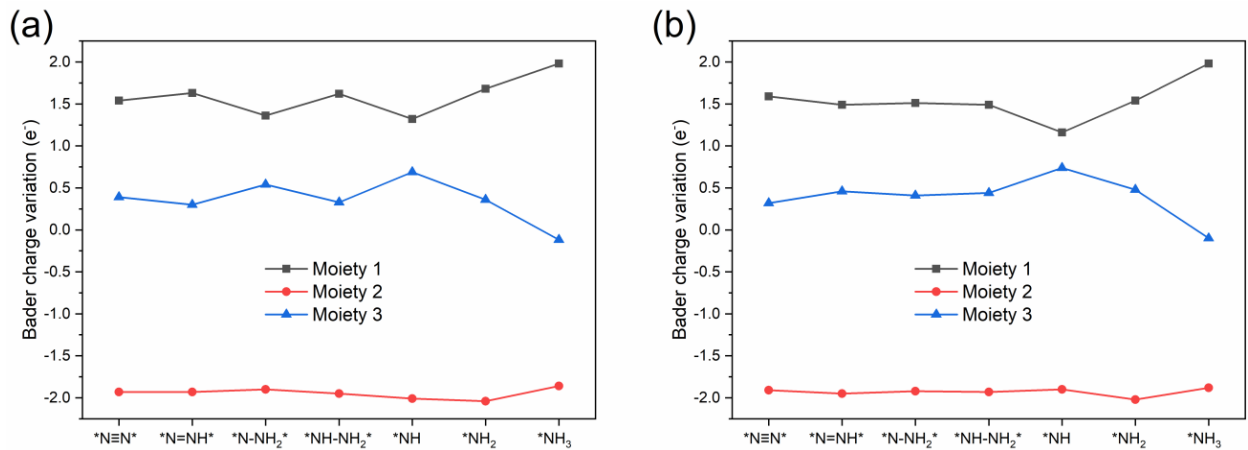

**Fig. 6** Bader charge variation of the three moieties along the most favorable NRR pathway for (a) Sc-TCNQ, and (b) Ti-TCNQ.

Since HER is the major competing reaction for NRR, we also compare the adsorption Gibbs free energy of $N_2$ and H for Sc, Ti, and V-TCNQ. Results shown in Fig. S4 indicate that the adsorption of $N_2$ is more energetically favorable than H, ensuring the high selectivity of NRR over HER.

### 3.3 Stability of TM-TCNQ Monolayers (TM=Sc and Ti)

In the final part, we test the stability of Sc and Ti-TCNQ, which are all good candidates for NRR electrocatalysts. First, to confirm that single metal aggregation will not occur on the surface of TCNQ, we compare the binding energy ($E_b$) of the single atom and the cohesive energy ($E_c$) for metal aggregation, which are defined as $E_b=E_{TM-TNCQ}-E_{TNCQ}-E_{TM}$ and $E_c=(E_{TM-bulk}-nE_{TM})/n$, respectively, where $E_{TM}$ and $E_{TM-bulk}$ represent the energy of single transition metal atom and total energy of transition metal in crystal, while n denotes the metal atoms in the unit cell of the crystal. $E_b$ values are calculated to be -9.29 and -9.98 eV for Sc and Ti on TCNQ, while the $E_c$ values are

-4.59 and -6.76 eV, indicating that binding to the substrate is more energetically favorable than aggregating into clusters for both Sc and Ti and the single atoms can be stably deposited onto the substrate as active sites.

To ensure the thermal stability of the SACs, we further perform AIMD simulations at 500 K and plot the total energy as a function of simulation time for Sc and Ti-TCNQ. Results in Fig. 7 exhibit that the total energy of Sc and Ti-TCNQ quickly converge and oscillate around the equilibrium. The structures after 5 ps AIMD simulations (inset in Fig. 7a and 7b) do not show an obvious structural reconstruction, and after geometric relaxation, the structures can recover again. These evidences clearly demonstrate that Sc and Ti-TCNQ are thermally stable with the temperature up to 500 K.

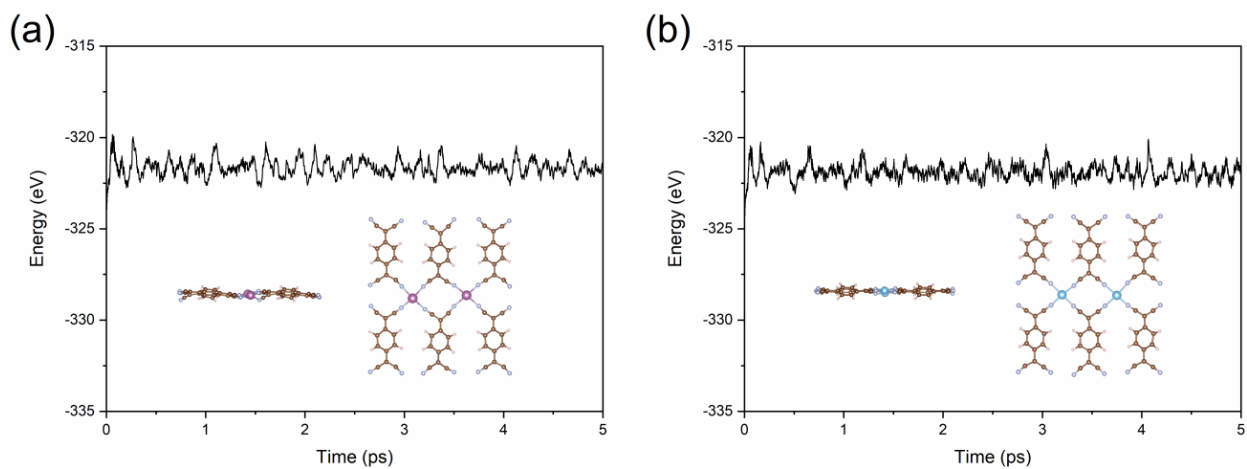

**Fig. 7** Total energy as a function of time during the AIMD simulations for (a) Sc-TCNQ and (b) Ti-TCNQ. Inset shows side and top view of the structures after 5 ps AIMD simulations, where C, N, H, Sc, and Ti are represented in brown, cyan, pale pink, purple, and blue, respectively.

**4. Conclusions**

In this work, by means of first-principles calculations, we systematically study the potential application of TM-TCNQ, a group of novel single-atom catalysts, on electrocatalytic NRR. From a two-step screening process, we identify three systems—Sc, Ti, and V-TCNQ with side-on $N_2$ adsorption pattern—from 34 structures. $N_2$ adsorption and activation are ensured by the unique electronic properties of TM-TCNQ and the 'acceptance-donation' mechanism. Gibbs free energy calculations prove that NRR Sc and Ti-TCNQ is thermodynamically favorable with low overpotential values of 0.33 and 0.22 V through the enzymatic-consecutive mixed pathway. In addition, NRR selectivity and stability of the catalysts are also confirmed. Overall, our work provides evidence that Sc and Ti-TCNQ can be promising NRR electrocatalysts. The findings and methodology in this work may be extended to the design of other single-atom catalysts.

AUTHOR INFORMATION

**Corresponding Author**

*E-mail: luox77@mail.sysu.edu.cn (X.L.); aphhuang@polyu.edu.hk (H.H.).

**Notes**

The authors declare no competing financial interests.

ACKNOWLEDGMENT

This work was supported by the Research Grants Council of the Hong Kong Special Administrative Region, China (Project No. PolyU152140/19E), the Hong Kong Polytechnic University (Project Nos. Q54V), and National Natural Science Foundation of China (No. 11804286), and the Fundamental Research Funds for the Central Universities (Grant No. 19lgpy263). The DFT calculations were partially performed on Apollo cluster at the Department of Applied Physics, the Hong Kong Polytechnic University.

# Supporting information

# Transition Metal-Tetracyanoquinodimethane Monolayers as Single-Atom Catalysts for Electrocatalytic Nitrogen Reduction Reaction

*Yiran Ying [a], Ke Fan [a], Xin Luo [b]\*, and Haitao Huang [a]\**

[a] Department of Applied Physics, The Hong Kong Polytechnic University, Hung Hom, Kowloon, Hong Kong, P.R. China

[b] School of Physics, Sun Yat-sen University, Guangzhou, Guangdong Province, P.R. China, 510275

**Corresponding Author**

*E-mail: luox77@mail.sysu.edu.cn; aphhuang@polyu.edu.hk

**Table S1.** Optimized lattice constants *a* and *b* for TM-TCNQ (TM=Sc-Zn, Mo, Ru-Pd, Ag, Pt, Au).

| System | *a* (Å) | *b* (Å) | System | *a* (Å) | *b* (Å) |
|---|---|---|---|---|---|
| **Sc-TCNQ** | 7.266 | 11.702 | **Zn-TCNQ** | 7.158 | 11.423 |
| **Ti-TCNQ** | 7.162 | 11.559 | **Mo-TCNQ** | 7.041 | 11.522 |
| **V-TCNQ** | 7.095 | 11.455 | **Ru-TCNQ** | 7.014 | 11.354 |
| **Cr-TCNQ** | 7.043 | 11.332 | **Rh-TCNQ** | 7.028 | 11.326 |
| **Mn-TCNQ** | 7.015 | 11.295 | **Pd-TCNQ** | 7.030 | 11.363 |
| **Fe-TCNQ** | 6.938 | 11.193 | **Ag-TCNQ** | 7.381 | 11.758 |
| **Co-TCNQ** | 6.886 | 11.161 | **Pt-TCNQ** | 7.022 | 11.368 |
| **Ni-TCNQ** | 6.974 | 11.208 | **Au-TCNQ** | 7.297 | 11.549 |
| **Cu-TCNQ** | 7.183 | 11.371 | | | |

**Table S2.** Adsorption energies $E_{ad}$ (eV) of $N_2$ onto the TM-TCNQ (TM=Sc-Zn, Mo, Ru-Pd, Ag, Pt, Au) with both end-on and side-on configurations. Catalysts with favorable $E_{ad}$ value for NRR are marked in red.

| System | End-on | Side-on | System | End-on | Side-on |
|---|---|---|---|---|---|
| **Sc-TCNQ** | -0.63 | -0.33 | **Zn-TCNQ** | -0.11 | -0.05 |
| **Ti-TCNQ** | -0.97 | -0.56 | **Mo-TCNQ** | -0.06 | -0.06 |
| **V-TCNQ** | -1.16 | -0.50 | **Ru-TCNQ** | -0.03 | -0.07 |
| **Cr-TCNQ** | -1.20 | -0.09 | **Rh-TCNQ** | -0.03 | -0.07 |
| **Mn-TCNQ** | -0.96 | -0.04 | **Pd-TCNQ** | -0.05 | -0.07 |
| **Fe-TCNQ** | -0.45 | -0.08 | **Ag-TCNQ** | -0.01 | -0.01 |
| **Co-TCNQ** | 0.08 | 0.04 | **Pt-TCNQ** | -0.05 | -0.07 |
| **Ni-TCNQ** | 0.05 | 0.02 | **Au-TCNQ** | -0.27 | -0.28 |
| **Cu-TCNQ** | -0.05 | -0.06 | | | |

**Table S3.** DFT-calculated total energy $E_{DFT}$, zero point energy $E_{ZPE}$, and entropic contribution term TS for NRR and HER intermediates on Sc-TCNQ.

| Adsorbates | $E_{DFT}$ (eV) | $E_{ZPE}$ (eV) | TS (eV) |
| --- | --- | --- | --- |
| * | -162.539 | 0 | 0 |
| *N-N | -179.776 | 0.177 | 0.141 |
| *N-NH | -181.070 | 0.399 | 0.125 |
| *N-N* | -179.468 | 0.161 | 0.145 |
| *N-NH* | -182.793 | 0.484 | 0.163 |
| *N-NH$_2$* | -186.253 | 0.474 | 0.281 |
| *NH-NH* | -186.306 | 0.719 | 0.146 |
| *N | -167.512 | 0.042 | 0.116 |
| *NH-NH$_2$* | -191.059 | 1.119 | 0.152 |
| *NH$_2$-NH$_2$* | -193.786 | 1.441 | 0.211 |
| *NH | -173.978 | 0.325 | 0.114 |
| *NH$_2$ | -179.883 | 0.641 | 0.139 |
| *NH$_3$ | -183.323 | 1.014 | 0.174 |
| *H | -166.185 | 0.145 | 0.027 |

**Table S4.** DFT-calculated total energy $E_{DFT}$, zero point energy $E_{ZPE}$, and entropic contribution term TS for NRR and HER intermediates on Ti-TCNQ.

| Adsorbates | $E_{DFT}$ (eV) | $E_{ZPE}$ (eV) | TS (eV) |
|---|---|---|---|
| * | -162.820 | 0 | 0 |
| *N-N | -180.394 | 0.217 | 0.089 |
| *N-NH | -182.135 | 0.404 | 0.060 |
| *N-N* | -179.972 | 0.204 | 0.096 |
| *N-NH* | -183.212 | 0.481 | 0.114 |
| *N-NH$_2$* | -187.236 | 0.826 | 0.104 |
| *NH-NH* | -186.476 | 0.711 | 0.147 |
| *N | -170.319 | 0.070 | 0.078 |
| *NH-NH$_2$* | -191.484 | 1.123 | 0.213 |
| *NH$_2$-NH$_2$* | -194.089 | 1.459 | 0.221 |
| *NH | -176.090 | 0.348 | 0.090 |
| *NH$_2$ | -180.398 | 0.659 | 0.115 |
| *NH$_3$ | -183.731 | 1.021 | 0.167 |
| *H | -166.752 | 0.162 | 0.021 |

**Table S5.** DFT-calculated total energy $E_{DFT}$, zero point energy $E_{ZPE}$, and entropic contribution term TS for NRR and HER intermediates on V-TCNQ.

| Adsorbates | $E_{DFT}$ (eV) | $E_{ZPE}$ (eV) | TS (eV) |
| --- | --- | --- | --- |
| * | -162.840 | 0 | 0 |
| *N-N | -180.601 | 0.216 | 0.133 |
| *N-NH | -182.660 | 0.424 | 0.104 |
| *N-N* | -179.932 | 0.177 | 0.117 |
| *N-NH* | -182.884 | 0.477 | 0.107 |
| *N-NH$_2$* | -187.238 | 0.825 | 0.097 |
| *NH-NH* | -186.794 | 0.770 | 0.110 |
| *N | -171.575 | 0.075 | 0.028 |
| *NH-NH$_2$* | -191.076 | 1.115 | 0.157 |
| *NH$_2$-NH$_2$* | -193.974 | 1.531 | 0.103 |
| *NH | -176.181 | 0.341 | 0.051 |
| *NH$_2$ | -180.154 | 0.677 | 0.095 |
| *NH$_3$ | -183.779 | 1.030 | 0.150 |
| *H | -166.756 | 0.173 | 0.021 |

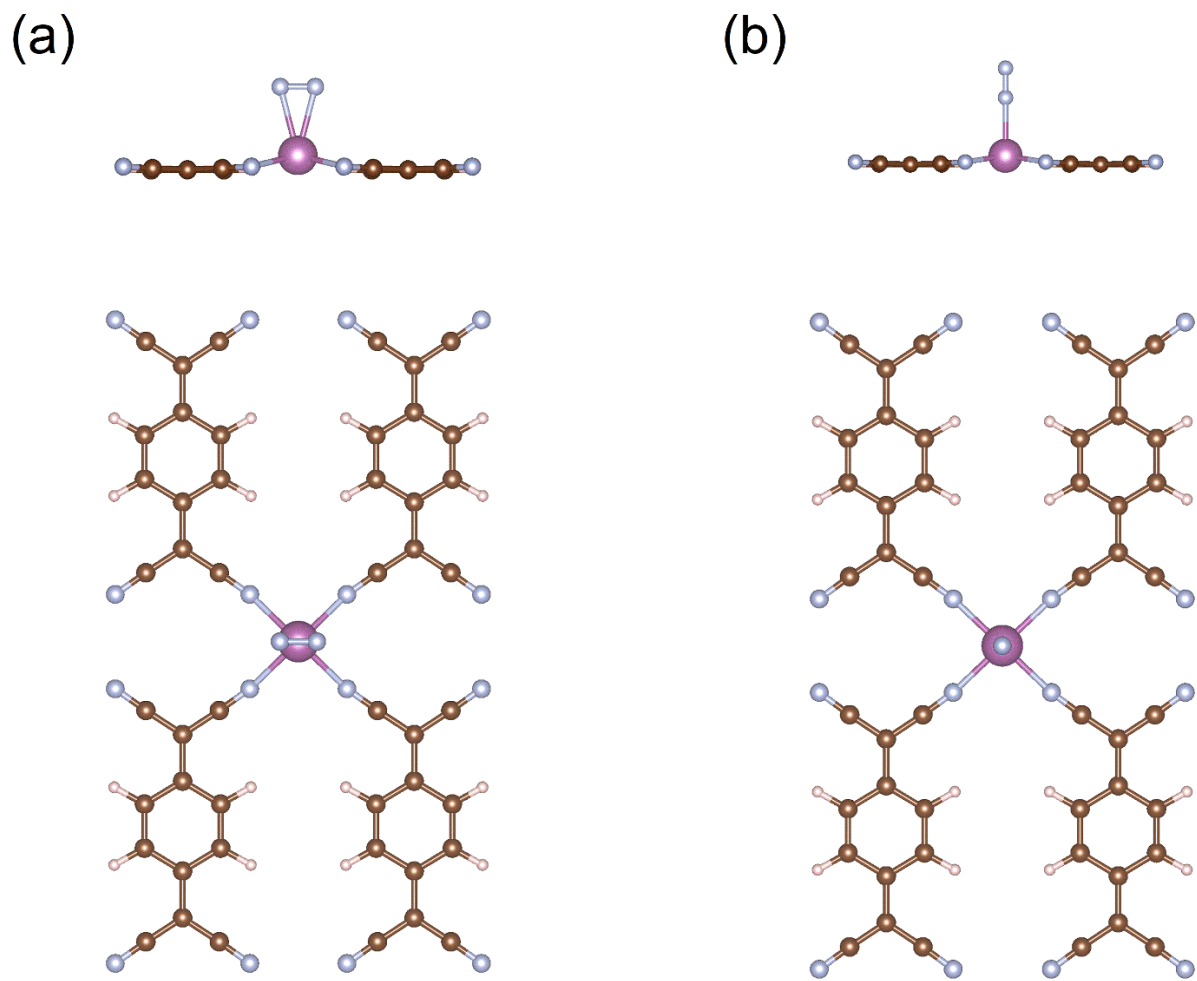

**Fig. S1** Side and top view of the schematic geometric structures of N$_2$ adsorbed on TM-TCNQ with (a) side-on configuration and (b) end-on configuration. TM, C, N, H are represented in purple, brown, cyan, and pale pink, respectively.

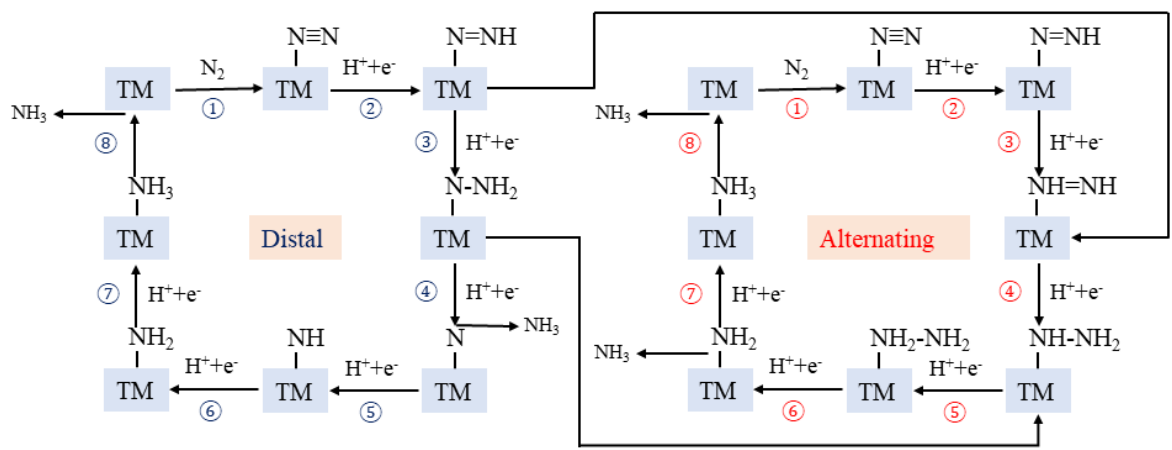

**Fig. S2** Schematic illustration of distal, alternating, and their mixed mechanisms for NRR on TM-TCNQ. TM denotes transition metal atom (active site) in the figures.

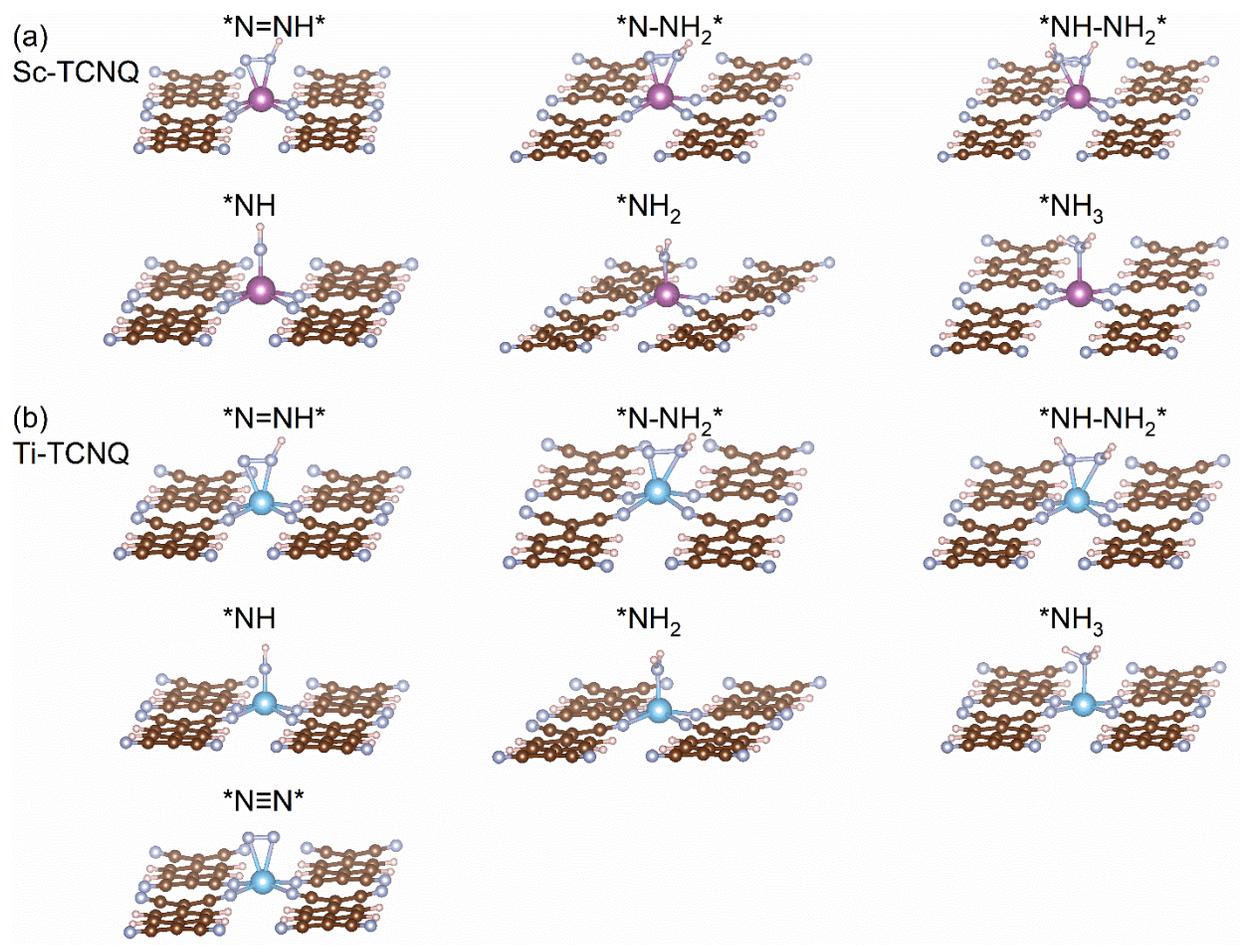

**Fig. S3** DFT-optimized structures for all intermediates in the NRR along the most favorable pathway for (a) Sc-TCNQ and (b) Ti-TCNQ except for *N≡N* on Sc-TCNQ (already shown in Fig. S1a). C, N, H, Sc, and Ti are represented in brown, cyan, pale pink, purple, and blue, respectively.

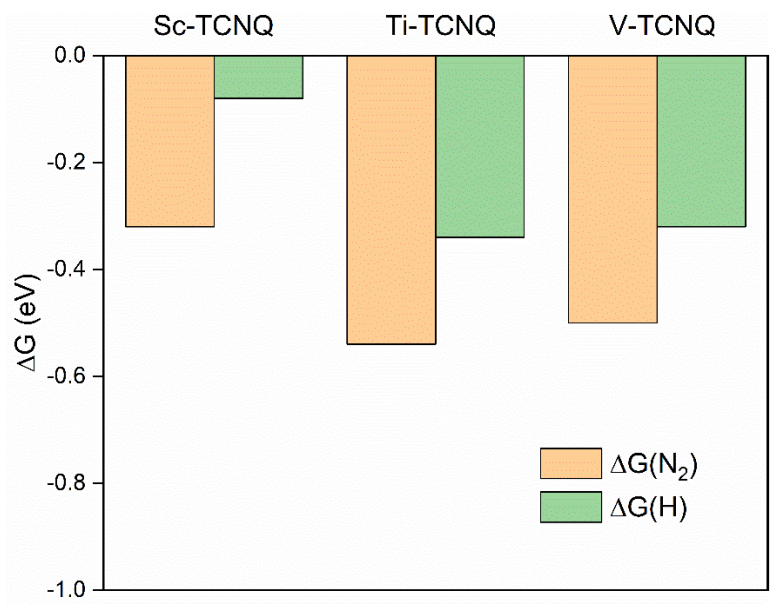

**Fig. S4** Comparison between adsorption Gibbs free energy for N₂ and H for Sc, Ti, V-TCNQ.